\documentclass{aa}  

\usepackage{graphicx}
\usepackage{txfonts}
\usepackage{multirow}
\usepackage{natbib}
\bibpunct{(}{)}{;}{a}{}{,} 
\usepackage[]{hyperref}
\usepackage{ulem}
\begin{document} 

    \titlerunning{Evidence for bipolar explosions in Type~IIP SNe}
    \authorrunning{T. Nagao et al.}

   \title{Evidence for bipolar explosions in Type~IIP supernovae}


   \author{T.~Nagao,
          \inst{1,2,3}
          K.~Maeda, \inst{4}
          S.~Mattila, \inst{1,5} 
          H.~Kuncarayakti, \inst{1,6} 
          M.~Kawabata, \inst{7}
          K.~Taguchi, \inst{4}
          T.~Nakaoka, \inst{8,9}
          A.~Cikota, \inst{10} 
          M.~Bulla, \inst{11,12,13}
          S.~S.~Vasylyev, \inst{14}\fnmsep\thanks{Steven Nelson Graduate Fellow} 
          C.~P.~Guti{\'e}rrez, \inst{15,16} 
          M.~Yamanaka, \inst{17}
          K.~Isogai, \inst{7,18}
          K.~Uno, \inst{4}
          M.~Ogawa, \inst{4}
          S.~Inutsuka, \inst{4}
          M.~Tsurumi, \inst{4}
          R.~Imazawa, \inst{9}
          \and
          K.~S.~Kawabata \inst{8,9}
          }

    \institute{
            Department of Physics and Astronomy, University of Turku, FI-20014 Turku, Finland
            \and
            Aalto University Mets\"ahovi Radio Observatory, Mets\"ahovintie 114, 02540 Kylm\"al\"a, Finland
            \and
            Aalto University Department of Electronics and Nanoengineering, P.O. BOX 15500, FI-00076 AALTO, Finland
            \and
            Department of Astronomy, Kyoto University, Kitashirakawa-Oiwake-cho, Sakyo-ku, Kyoto 606-8502, Japan
            \and
            School of Sciences, European University Cyprus, Diogenes Street, Engomi, 1516, Nicosia, Cyprus
            \and
            Finnish Centre for Astronomy with ESO (FINCA), University of Turku, FI-20014, Finland
            \and
            Okayama Observatory, Kyoto University, 3037-5 Honjo, Kamogatacho, Asakuchi, Okayama 719-0232, Japan
            \and
            Hiroshima Astrophysical Science Center, Hiroshima University, Kagamiyama 1-3-1, Higashi-Hiroshima , Hiroshima 739-8526, Japan
            \and
            Department of Physical Science, Hiroshima University, Kagamiyama 1-3-1, Higashi-Hiroshima 739-8526, Japan
            \and
            Gemini Observatory / NSF's NOIRLab, Casilla 603, La Serena, Chile
            \and
            Department of Physics and Earth Science, University of Ferrara, via Saragat 1, I-44122 Ferrara, Italy
            \and
            INFN, Sezione di Ferrara, via Saragat 1, I-44122 Ferrara, Italy
            \and
            INAF, Osservatorio Astronomico d'Abruzzo, via Mentore Maggini snc, 64100 Teramo, Italy
            \and
            Department of Astronomy, University of California, Berkeley, CA 94720-3411, USA
            \and
            Institut d'Estudis Espacials de Catalunya (IEEC), Edifici RDIT, Campus UPC, 08860 Castelldefels (Barcelona), Spain
            \and
            Institute of Space Sciences (ICE, CSIC), Campus UAB, Carrer de Can Magrans, s/n, E-08193 Barcelona, Spain
            \and
            Amanogawa Galaxy Astronomy Research Center (AGARC), Graduate School of Science and Engineering, Kagoshima University, 1-21-35 Korimoto, Kagoshima, Kagoshima 890-0065, Japan
            \and
            Department of Multi-Disciplinary Sciences, Graduate School of Arts and Sciences, The University of Tokyo, 3-8-1 Komaba, Meguro, Tokyo 153-8902, Japan
             }

   \date{Received ***; accepted ***}

 
  \abstract
   {}
   {Recent observations of core-collapse supernovae (SNe) suggest aspherical explosions. Globally aspherical structures in SN explosions are regarded as the key for understanding their explosion mechanism. However, the exact explosion geometries from the inner cores to the outer envelopes are poorly understood.}
   {Here, we present photometric, spectroscopic and polarimetric observations of the Type~IIP SN~2021yja and discuss its explosion geometry, in comparison to those of other Type~IIP SNe that show large-scale aspherical structures in their hydrogen envelopes (SNe 2012aw, 2013ej and 2017gmr).}
   {During the plateau phase, SNe 2012aw and 2021yja exhibit high continuum polarization characterized by two components with perpendicular polarization angles. This behavior can be interpreted to be due to a bipolar explosion, composed of a polar (energetic) and an equatorial (bulk) components of the SN ejecta. In such a bipolar explosion, an aspherical axis created by the polar ejecta would be dominating at early phases, while the perpendicular axis along the equatorial ejecta would emerge at late phases after the receding of the photosphere in the polar ejecta. The interpretation of the bipolar explosions in SNe~2012aw and 2021yja is also supported by other observational properties, including the time evolution of the line velocities and the line shapes in the nebular spectra. The polarization of other Type~IIP SNe that show large-scale aspherical structures in the hydrogen envelope (SNe 2013ej and 2017gmr) is also consistent with the bipolar-explosion scenario, although this is not conclusive.
   }
   {}


   \maketitle
%

\section{Introduction} \label{sec:introduction}

Core-collapse supernovae (SNe) are catastrophic explosions due to the neutrino heating from a newly-created proto-neutron star after a collapse of the iron core \citep[the neutrino-driven mechanism; e.g.,][]{Janka2017}, characterizing the deaths of massive stars. For a long time, it has been challenging to conduct first-principles simulations of SN explosions due to the limitation of the computational power. There are many complications and difficulties concerning nuclear and neutrino physics, the three-dimensional progenitor structures, and numerical challenges in the treatments of the transfer of the radiation and neutrinos and in general relativistic three-dimensional hydrodynamics calculations \citep[e.g.,][]{Burrows2021}. Recent advancements in the computational power and the methodology for the calculations are gradually allowing successful cases of SN explosions \citep[e.g.,][]{Lentz2015,Melson2015,Radice2017,Vartanyan2018,Burrows2020}. However, such successful explosions are limited to a small number of parameter sets of massive stars, where the majority of massive stars are still difficult to be exploded as realistic SNe in a computer. Therefore, we are missing some key elements in our understanding of the SN explosions.

It has been suggested that some enhancements of multi-dimensional hydrodynamic instabilities, such as convective motion, the standing-accretion-shock instability \citep[][]{Blondin2003} and/or Rayleigh-Taylor instability, might play a crucial role because multi-dimensional motions of the gas can increase the neutrino heating efficiency in the gain region \citep[e.g.,][]{Burrows2020}. In fact, using state-of-the-art three-dimensional simulations based on the neutrino-driven mechanism, \citet[][]{Burrows2024} have demonstrated that the successful explosions generally have global asymmetric structures and the asymmetry is correlated with the explosion energy.
This is also suggested by recent observations of core-collapse SNe. Polarimetric observations have revealed large-scale aspherical structures in the ejecta of hydrogen-rich (Type~II) SNe. 
Historically, a low level of polarization ($\sim 0.1$\%) has been measured during the plateau phase, followed by a rapid increase of continuum polarization at the beginning of the tail phase \citep[e.g.,][]{Leonard2001,Leonard2006,Chornock2010,Kumar2016}. This polarimetric feature can be explained by an asymmetric helium core being revealed when the outer hydrogen envelope becomes optically thin as indicated by the light curve falling off the plateau. Recently, \citet[][]{Nagao2019} found an unprecedentedly highly-extended aspherical explosion in Type~II SN~2017gmr indicated by an early rise of polarization even during the plateau phase. This indicates that asymmetries are present not only in the helium core but also in a substantial portion of the hydrogen envelope. Furthermore, \citet[][]{Nagao2024}, using a large sample of SNe, discovered a correlation between the extension of the aspherical structures and the explosion energy, implying that development of a global aspherical structure might be a key element in the SN explosions. Such global aspherical structures in core-collapse SNe are also required by statistical analysis of line shapes in the nebular spectra of stripped-envelope SNe \citep[e.g.,][]{Fang2024}. This study demonstrated that the spatial distributions of the oxygen-burning ash and the unburnt oxygen follow the configurations of bipolar explosions. Moreover, they found that the degree of asphericity, in such bipolar explosions, increased towards explosions from heavier progenitors.

In this paper, we present photometric, spectroscopic and polarimetric observations of the Type~IIP SN~2021yja and discuss its explosion geometry together with those of other Type~IIP SNe that show large-scale aspherical structures in the hydrogen envelope \citep[SNe 2012aw, 2013ej and 2017gmr;][]{Nagao2024}.
SN~2021yja was discovered by the Asteroid Terrestrial-impact Last Alert System \citep[][]{Smith2020} on 8.55 September 2021 UT \citep[59465.55 MJD;][]{Tonry2021}. It is located in NGC~1325 (Galaxy Morphology: SA(s)bc; from NED\footnote{NASA/IPAC Extragalactic Database}) at $z=0.005307 \pm 0.000005$ and receding with a velocity of $v_{\rm{gal}}=1591\pm1$ km s$^{-1}$ \citep[][]{Springob2005}. The object was not detected on 6.48 September 2021 UT \citep[59463.48 MJD;][]{Tonry2021}. We adopt the explosion date to be MJD 59464.40, estimated by \citet[][]{Hosseinzadeh2022}, which is about the middle point between the last non-detection and the discovery dates. The object was classified as a Type~II SN about a day after the discovery \citep[59466.60 MJD;][]{Pellegrino2021}, and showed prototypical observational properties for Type IIP SNe \citep[e.g.,][]{Vasylyev2022,Hosseinzadeh2022}. 
We adopted the Milky Way and host galaxy reddening of $E(B-V)_{\rm{MW}}=0.0191$ mag \citep[][]{Schlafly2011} and $E(B-V)_{\rm{host}}=0.085$ mag derived from the Na~I~D absorption lines by \citet[][]{Hosseinzadeh2022}, respectively, and thus a total reddening of $E(B-V)=0.104$ mag. For the extinction correction, we adopt the extinction law by \citet[][]{Fitzpatrick1999}. We adopt the distance and distance modulus for the SN of $23.4^{+5.4}_{-4.4}$ Mpc and $\mu=31.85 \pm 0.45$ mag, respectively, which were derived by the Tully-Fisher estimate \citep[][]{Tully2016}. In this work, we estimated the timing of the end of the plateau phase as the mid-point of the luminosity drop from the plateau phase to the tail phase following \citet[][]{Nagao2024}, by fitting the time evolution of the $g$-band light curve
taken from \citet[][]{Hosseinzadeh2022} with an artificial function \citep[Equation~1 in][]{Nagao2024}: $t=59590.57$ MJD.

\section{Observations} \label{sec:observation}

We performed photometric, spectroscopic and polarimetric observations of the Type~IIP SN~2021yja. The details on the photometric and spectroscopic observations are described in Appendix~\ref{sec:app_photo} and \ref{sec:app_spec}. 
We obtained spectro- and imaging polarimetry of the Type IIP SN 2021yja, using the FOcal Reducer/low-dispersion Spectrograph 2 (hereafter FORS2) mounted at the Cassegrain focus of the Very Large Telescope (VLT) UT1 telescope at the Paranal observatory and the Alhambra Faint Object Spectrograph and Camera (ALFOSC) mounted on the 2.56-m Nordic Optical Telescope (NOT) at the Roque de los Muchachos Observatory. The observing logs are shown in Tables~\ref{tab:spec_pol} and \ref{tab:image_pol}. 
It it noted that spectropolarimetric observations for SN~2021yja have also been conducted and reported by \citet[][]{Vasylyev2024}. Here, our polarimetric data cover later epochs.

The spectropolarimetric observations and their analysis are similar as reported in \citet[][]{Nagao2023}. We observed SN~2021yja using the low-resolution G300V grism and a half-wave retarder plate (HWP) with the optimal set of HWP angles of $0^{\circ}$, $22.5^{\circ}$, $45^{\circ}$ and $67.5^{\circ}$. The raw data were reduced with IRAF \citep[][]{Tody1986,Tody1993} using standard methods as described, e.g., in \citet[][]{Patat2006}. From the reduced object frames, we extracted the ordinary and extraordinary beams of the SN with a fixed aperture size of $10$ pixels, and rebinned the extracted spectra to 50 Angstrom bins to allow a better signal-to-noise ratio. We also corrected the observations for HWP zeropoint angle chromatism, using tabulated values for the zero-angle given in the FORS2 user manual \footnote{\url{http://www.eso.org/sci/facilities/paranal/instruments/fors/doc/VLT-MAN-ESO-13100-1543_P07.pdf}}. The wavelength scale was corrected to the rest frame using the host-galaxy redshift. For deriving the continuum polarization from the spectropolarimetric data, we used the wavelength ranges between 6800 and 7200 {\AA} and between 7820 and 8140 {\AA}, following \citep[][]{Nagao2024}.
For the imaging polarimetry, the same instrumental set-up was adopted, using a narrow-band filter (FILT\_815\_13) with the VLT and two broad band filters ($V$ and $R$) with the NOT instead of the grism in the optical path. We applied bias subtraction and flat-field correction to all the frames and then performed aperture photometry on each detected source. For the aperture photometry, we used an aperture size that is twice as large as the full-width-half-maximum (FWHM) of the ordinary beam’s point-spread function and a background region whose inner and outer radii are twice and four times as large as the FWHM, respectively. Based on these values, we calculated the Stokes parameters, the linear polarization degree and the polarization angle. When calculating the polarization degrees, we subtracted the polarization bias, following \citet[][]{Wang1997}.
Since the polarization degrees on the H$\alpha$ emission line, where the intrinsic polarization is supposed to be depolarized and thus the interstellar polarization (ISP) to be dominant \citep[e.g.,][]{Nagao2023}, are close to zero, we assume that the ISP in SN~2021yja is negligibly small, which is also adopted by \citet[][]{Vasylyev2024}.

\section{Results and discussion} \label{sec:result}

   \begin{figure*}
   \centering
            \includegraphics[width=\hsize]{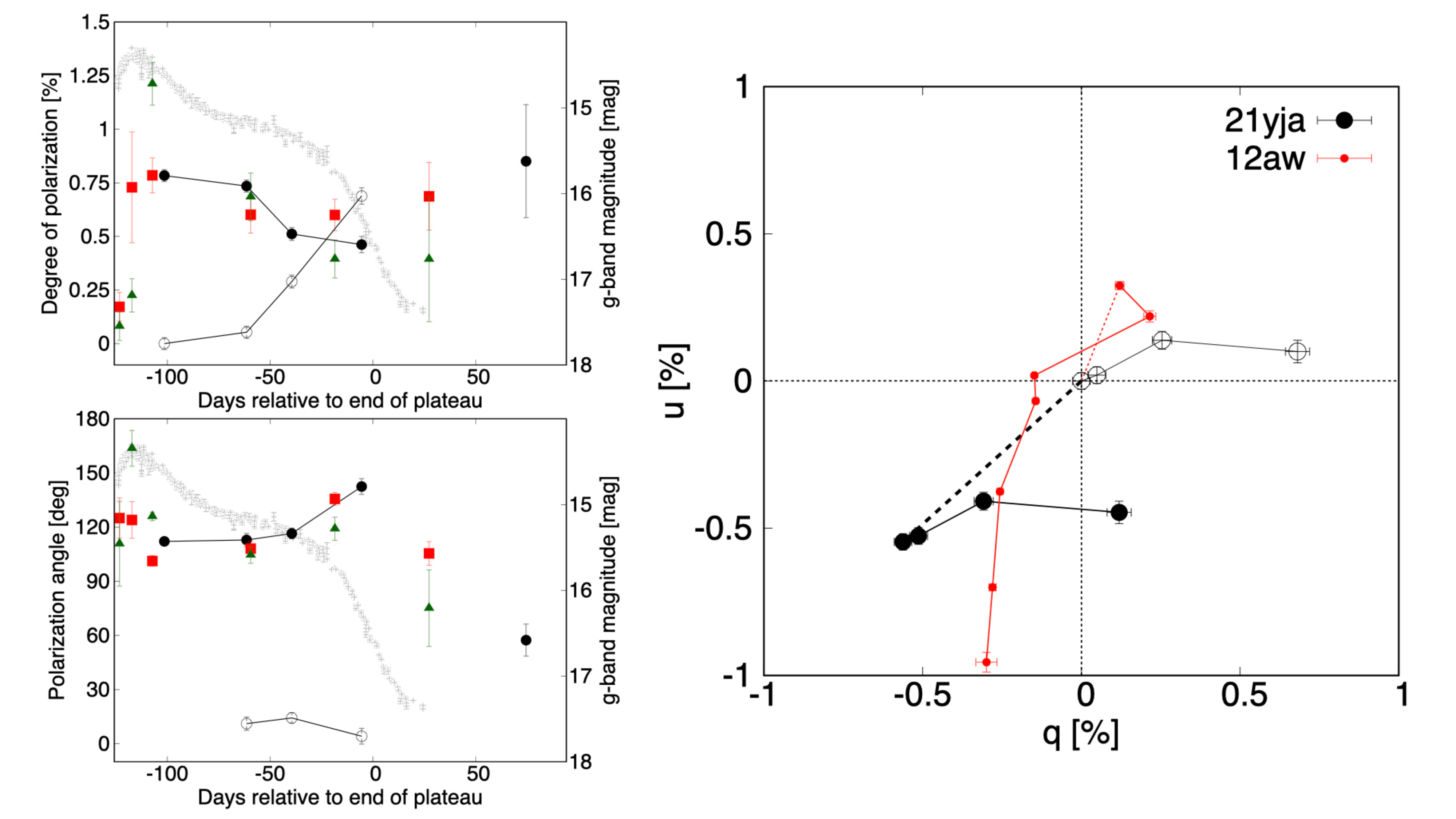}
      \caption{
      Left panel: Polarization degree and angle of SN~2021yja. The filled black circles connected by a solid line show the continuum polarization derived from the polarization spectra, while the filled green triangles and red squares show the values from the $V$- and $R$-band imaging polarimetry. The open black circles represent only the second component of the continuum polarization. The gray crosses trace the $g$-band light curve of SN 2021yja taken from \citet[][]{Hosseinzadeh2022}. Right panel: Time evolution of the continuum polarization in the q-u plane, compared to that of SN~2012aw. The first-epochs are indicated with dotted lines. The open black circles are the same in the left panels, representing only the second component of the continuum polarization.
              }
      \label{fig:fig1}
   \end{figure*}

Figure~\ref{fig:fig1} shows the continuum polarization of SN~2021yja, derived from spectropolarimetry and from the narrow/broad-band polarization. As also seen in \citet[][]{Vasylyev2024}, SN~2021yja shows a distinct evolution of the continuum polarization compared with the majority of other Type~IIP SNe. Most Type~IIP SNe show low polarization degrees at early phases and a sudden rise to $\sim 1$\% at a certain point in the plateau phase with a constant polarization angle \citep[see, e.g.,][]{Nagao2024}. In contrast, SN~2021yja shows a high degree of polarization ($\sim 0.8$\%) at early phases and a decline of the polarization degree with time as well as a gradual change of the polarization angle. This implies multi-axial structures in the hydrogen envelope, whose relative contributions to the polarization evolve with time. The $V$- and $R$-band data are generally consistent with the continuum polarization estimated from the polarization spectra, even though they should be affected by line depolarization/polarization (see the polarization spectra in Appendix~\ref{sec:app_pol}) and have larger errors. Hereafter, when discussing the continuum polarization of SN~2021yja, we only use the values derived from the polarization spectra. It is noted that the polarization angles in the tail phase estimated from the narrow-band, $V$- and $R$-band imaging polarimetry are not stable and thus might indeed be affected by line polarization. In the tail phase, the line polarization can dominate over the continuum polarization \citep[see, e.g., the cases of SNe~2017gmr and 2013ej;][]{Nagao2019,Nagao2021}. Since we do not have spectropolarimetry for the tail phase, we do not further discuss the polarization in the tail phase in this paper.

We adopt the continuum polarization derived from the first-epoch spectropolarimetric data as the ``first component" ($P=0.78$ \%, $\theta=112.1$ degrees). Then, we subtract the first component from the original continuum polarization. The remaining polarization shows a relatively constant polarization angle ($\theta \sim 15$ degrees), which we call the ``second component". This means that the continuum polarization during the plateau phase can be explained by these two components. 
It is noted that, if we assume the polarization degree of the first component with $\theta=112.1$ degrees is not constant but variable in time, the remaining polarization after the subtraction of the first component from the original polarization still shows time-variable polarization angles. Therefore, in order to remove the time variability of the polarization angle of the remaining polarization, at least two more additional polarization sources would be required. Consequently, we need at least three components of the polarization in total as the origin of the continuum polarization.
In addition, any of these components would not show typical time evolution of the polarization degrees as in other Type~IIP SNe \citep[][]{Nagao2024}.
The difference between the polarizatiton angles of these components is $\sim 95$ degrees. In the q-u plane, the data points of the second component (open black circles) are clearly located nearly to the opposite direction relative to the first component (dotted line) towards the origin of the q-u plane, demonstrating that the axes of the aspherical structures implied by the first and second components are almost perpendicular (Figure~\ref{fig:fig1}).

In the literature, several other Type~II SNe have been reported to clearly show time-variable polarization angles in their polarization, implying multi-axis structures in the ejecta \citep[SNe~2012aw, 2013ej and 2023ixf;][]{Nagao2021,Nagao2024,Vasylyev2024}. 
SN~2023ixf showed high continuum polarization with a different polarization angle from that for later phases only within a few days after the explosion, which coincides with the existence of highly ionized ``flash" features in the spectra \citep[][]{Vasylyev2022}. Since the flash features are created by highly ionized gas in confined circumstellar material (CSM) due to high energy photons created in an interaction between the SN ejecta and the CSM\citep[e.g.,][]{Boian2020}, this short-lived component of the polarization is interpreted to reflect the aspherical structure not of the SN ejecta but of the confined CSM
\citep[][]{Vasylyev2024}.
In the case of SN~2013ej, the evolution of the polarization angle happened using the majority part of the plateau phase \citep[][]{Nagao2021}. At the same time, the SN showed observational signs of a major CSM interaction at least during first several tens days \citep[e.g.,][]{Bose2015}. This evolution of the polarization is understood as the change in the contributions from two different aspherical structures originated from both an aspherical CSM interaction and aspherical SN ejecta. 
On the contrary, SNe~2012aw and 2021yja do not show observational features from a strong CSM interaction that could change the global properties of the SN ejecta, although a weak CSM interaction is predicted \citep[e.g.,][]{Bose2013,Hosseinzadeh2022,Vasylyev2022}. The low degree of the early-phase broad-band polarization in SN~2021yja also disagrees with the scenario of a major spherical CSM interaction as in the case of SN~2013ej. Furthermore, the change in the polarization angle between the first and second components in the cases of SNe~2012aw and 2021yja is close to $\sim 90$ degrees, demonstrating that the axes of these two components are close to perpendicular, alike in the case of SN~2013ej ($\sim 20$ degrees; Nagao et al. 2021). If the first and second components originate from an aspherical CSM interaction and aspherical SN ejecta (as in the case of SN~2013ej), the angle change should be random due to the random orientation between the CSM structure and the aspherical SN explosion, unless there is some mechanism to create a common preferential axis for the mass loss and the SN explosion.

   \begin{figure}
   \centering
            \includegraphics[width=\hsize]{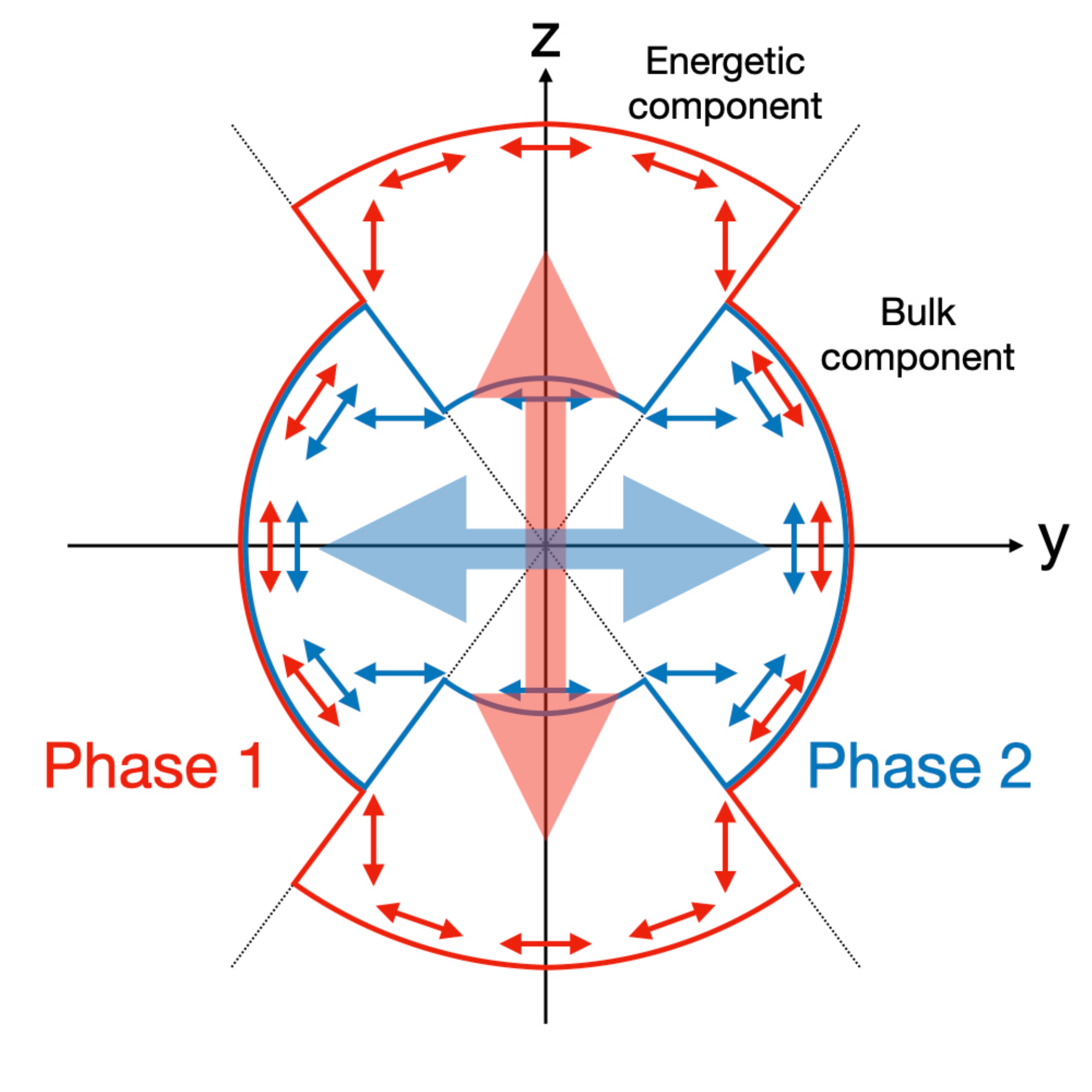}
      \caption{
      Schematic presentation of the continuum polarization in SN2021yja (observer's view) created by a bipolar explosion. At early phases, the photosphere is extended vertically due to the energetic ejecta component, which creates the first vertical polarization component (Phase~1).
      Then, the shape of the photosphere is crushed horizontally due to the receding of the photosphere in the energetic component, which creates the second horizontal polarization component (Phase~2). This explains the rotation of the observed polarization angle.
      The polarization vectors of important individual photons from the photosphere are indicated with small arrows. The big arrows show the net polarization vectors of all the photons.
              }
      \label{fig:fig2}
   \end{figure}

We propose that this is evidence of the presence for two-component ejecta driven by a bipolar explosion (a collimated energetic component and a canonical bulk SN component), as implied for the explanation of the polarization in SN~2012aw by \citet[][]{Nagao2024}. We consider a similar situation with the bipolar ejecta model proposed by \citet[][]{Kaplan2020}. The schematic picture of this scenario is shown in Figure~\ref{fig:fig2}. If an energy injection is stronger towards the polar directions than to the equatorial directions, the ejecta in the polar direction (the energetic component) would create a larger photosphere than that in the equatorial directions (the bulk component) at early phases (Phase~1). Since the receding of the photosphere would happen earlier in the polar directions than in the equatorial directions, the size of the photosphere in the energetic component becomes smaller than that in the bulk component late in the plateau phase (Phase ~2). This transition from Phase~1 to Phase~2 would create the 90-degree change in the angle of the axis of the aspherical structures. Such morphology change has also been predicted by past radiative transfer calculations in a bipolar explosion \citep[e.g.,][]{Maeda2006,Dessart2024}.

The relative strength of the first and second components and its time evolution should depend on the ejecta parameters (mass, energy, collimation degree) for the polar and equatorial directions and on the viewing angle. In particular, the timing of the transition from Phase~1 to Phase~2 should not strongly depend on the viewing angle, but on the explosion parameters and the degree of the collimation of the energetic component. For example, if we assume that the explosion properties are independent for these perpendicular directions and that only the inputted specific explosion energy to these components is different, the timing of the transition from Phase~1 to Phase~2 can be roughly estimated from the relative explosion energy using the scaling relation in a simple one-zone model of homogeneously expanding gas with radiative diffusion proposed by 
\citet[][]{Kasen2009}: $t_{\rm{sn}} \propto E^{-1/6}$. In the case that the energetic component has 10 times larger specific energy than the bulk component, this transition would happen at 70\% of the plateau length of the bulk component, although it should be a relatively smooth transition in reality. In fact, the timings of this phase transition in SNe~2012aw and 2021yja are different. There was an earlier transition in the case of SN~2012aw compared to SN~2021yja. This might imply that there is diversity in the bipolar structures in Type~IIP SNe. It is also noted that, since the relative strength of the two components depends on the viewing angle, the different viewing angles might also contribute to the different timing of the exchange of these two components to some extent.

There may exist other possibilities to create such a bipolar structure before the explosion, e.g., by some unknown aspherical energy injection just before the explosion or by extremely rapid rotation. However, SN~2021yja shows low polarization at very early phases ($\lesssim 0.2$ \% at Phase +2.78), which might suggest that the outermost layer of the ejecta is relatively spherical. Therefore, this bipolar structure in SN~2021yja might not be due to the inherent aspherical structure in the progenitor star but a bipolar explosion.
It is noted that the low polarization at early phases might be achieved due to the high optical depth for electron scattering in the outermost layer \citep[e.g.,][]{Dessart2011}, even if the progenitor originally has an aspherically-extended envelope.
This effect depends on the density distribution in the extended envelope and should be quantitatively tested.

The other Type~IIP SN that show large-scale aspherical structures in the hydrogen envelope \citep[SN~2017gmr][]{Nagao2019} did not show the two-component polarization, which is a strong support for a bipolar explosion, as in the cases of SNe~2012aw and 2021yja. However, the earliest polarimetric observation of SN~2017gmr was conducted 45.40 days after the first detection, and thus the first component might have already been covered by the second component \citep[see the case of SN~2012aw in][]{Nagao2024}. Alternatively, the first component might be weak due to the viewing angle. So far, all the Type~IIP SNe that show global aspherical structures in the hydrogen envelope \citep[energetic Type~II SNe][]{Nagao2024} are consistent with the bipolar-explosion scenario. It is important to increase the sample size to probe the diversity in the bipolar explosions.


   \begin{figure*}
            \includegraphics[width=0.5\hsize]{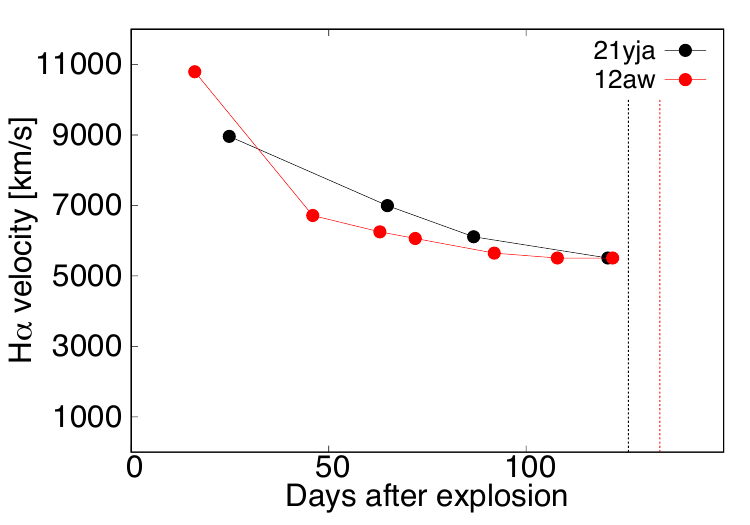}
            \includegraphics[width=0.5\hsize]{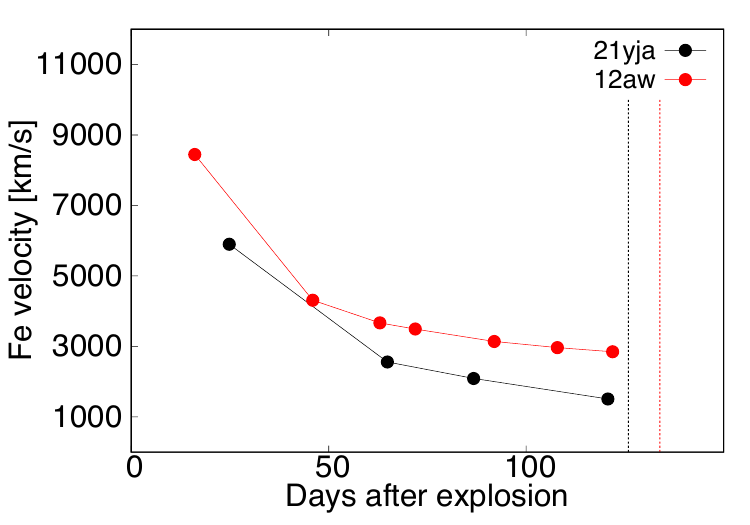}
      \caption{Time evolution of the H$\alpha$ and Fe~II velocities, derived from the absorption minima of the respective lines. The vertical dotted lines indicate the timings of the end of the plateau phase.
              }
      \label{fig:fig3}
   \end{figure*}

The interpretation of the bipolar explosions in SNe~2012aw and 2021yja are also supported by other observational properties. The general photometric and spectroscopic properties of SNe~2012aw and 2021yja are relatively similar (e.g., plateau luminosity, plateau length, the velocity evolution of the H$\alpha$ line; see Appendix~\ref{sec:app_photo} and \ref{sec:app_spec}), implying that the basic parameters of the SN ejecta are relatively similar. However, they show a different feature in the velocity evolution of the Fe~II $\lambda$5169 line, which is often used as a tracer of the velocity of the photosphere. At early phases, both SNe show similar velocity evolution for the Fe~II line as for the H$\alpha$ line, while SN~2021yja shows slower velocities than SN~2012aw at later plateau phase alike in the H$\alpha$ line (Figure~\ref{fig:fig3}). This observational property might be explained with different viewing angles towards relatively similar bipolar explosions. The viewing angle of SN~2021yja is closer to the polar direction compared to that of SN~2012aw. Thus, the slowdown of the Fe~II velocity, i.e., the photospheric velocity, might be related to the receding of the photosphere in the energetic component of the ejecta, i.e., the transition from Phase~1 to Phase~2. This might also be the reason why the relative strength of the first polarization component compared to the second component in SN~2021yja is larger than that in SN~2012aw.


\begin{figure}
\centering
            \includegraphics[width=\hsize]{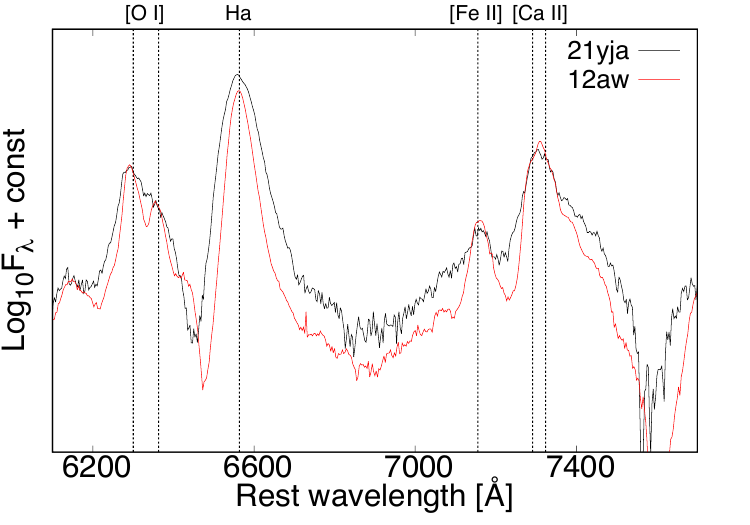}
      \caption{
        ALFOSC/NOT spectrum of SN~2021yja at 352.82 days after the explosion, compared to the spectrum of SN~2012aw obtained with the same instrument and telescope (ALFOSC/NOT) at +369 days after the explosion \citep[][]{Jerkstrand2014}. Wavelengths of prominent lines are indicated with vertical dashed lines.
              }
      \label{fig:fig4}
   \end{figure}

Nebular spectroscopy is also a powerful tool to study the geometry of the SN ejecta, especially for the inner parts 
\citep[e.g.,][]{Maeda2008,Taubenberger2009,Fang2024}, while polarimetry during the plateau phase provides information on the structure of the hydrogen envelope as discussed above. Figure~\ref{fig:fig4} shows a part of the nebular spectra of SNe~2021yja and 2012aw, taken with the ALFOSC/NOT (see Appendix~\ref{sec:app_spec} for details). SN~2021yja shows broader [Fe II] $\lambda$ 7155 line than that observed in SN~2012aw, although the Fe~II velocity during the plateau phase in SN~2021yja was slower than that seen in SN~2012aw. In addition, the [Ca II] $\lambda\lambda$ 7291,7323 lines are wider than those seen in SN~2012aw, while their [O I] $\lambda\lambda$ 6300,6363 lines are similar. These properties might also be explained by a scenario with the viewing angle effects of a bipolar explosion: The viewing angle of SN~2021yja is closer to the polar direction compared to that of SN~2012aw. Since the energetic ejecta component creates more burning ash (Fe and Ca) than the bulk component, the relative line widths of the iron and calcium lines to the oxygen line can be larger, if the viewing angle is closer to the polar direction. This interpretation is consistent with the properties of the polarization and the line velocities. This also suggests that the aspherical structure of the hydrogen envelope extends into the inner core, maintaining its shape. This provides strong evidence in favour of the bipolar explosion scenario, disfavouring the possibility of the CSM interaction scenario.

\begin{acknowledgements}

This work is partly based on observations collected at the European Organisation for Astronomical Research in the Southern Hemisphere (ESO) under programmes 108.228k.001 and 108.228k.002.
This work is partly based on observations made under program IDs P63-016, P64-023 and P65-005 with the Nordic Optical Telescope, owned in collaboration by the University of Turku and Aarhus University, and operated jointly by Aarhus University, the University of Turku and the University of Oslo, representing Denmark, Finland and Norway, the University of Iceland and Stockholm University at the Observatorio del Roque de los Muchachos, La Palma, Spain, of the Instituto de Astrofisica de Canarias.
The data from the Seimei and Kanata telescopes were taken under the KASTOR (Kanata And Seimei Transient Observation Regime) project (Seimei program IDs: 21B-N-CT09, 21B-K-0004, 21B-K-0011, 22A-K-0004, 22A-O-0003, 22B-N-CT10, 22B-K-0003). The Seimei telescope at the Okayama Observatory is jointly operated by Kyoto University and the Astronomical Observatory of Japan (NAOJ), with assistance provided by the Optical and Near-Infrared Astronomy Inter-University Cooperation Program.
This research was made possible through the use of the AAVSO Photometric All-Sky Survey (APASS), funded by the Robert Martin Ayers Sciences Fund and NSF AST-1412587.
TN acknowledges support from the Research Council of Finland projects 324504, 328898 and 353019.
KM acknowledges support from the Japan Society for the Promotion of Science (JSPS) KAKENHI grant (JP20H00174 and JP24H01810) and by the JSPS Open Partnership Bilateral Joint Research Project between Japan and Finland (JPJSBP120229923).
SM was funded by the Research Council of Finland project 350458.
HK was funded by the Research Council of Finland projects 324504, 328898, and 353019.
CPG acknowledges financial support from the Secretary of Universities and Research (Government of Catalonia) and by the Horizon 2020 Research and Innovation Programme of the European Union under the Marie Sk\l{}odowska-Curie and the Beatriu de Pin\'os 2021 BP 00168 programme, from the Spanish Ministerio de Ciencia e Innovaci\'on (MCIN) and the Agencia Estatal de Investigaci\'on (AEI) 10.13039/501100011033 under the PID2020-115253GA-I00 HOSTFLOWS project, and the program Unidad de Excelencia Mar\'ia de Maeztu CEX2020-001058-M.
This work was supported by JST, the establishment of university fellowships towards the creation of science technology innovation, Grant Number JPMJFS2129.
\end{acknowledgements}

%
%

\bibliographystyle{aa} 
\bibliography{aa.bib}

\appendix

\section{Polarimetry} \label{sec:app_pol}

\begin{figure*}
   \centering
            \includegraphics[width=\hsize]{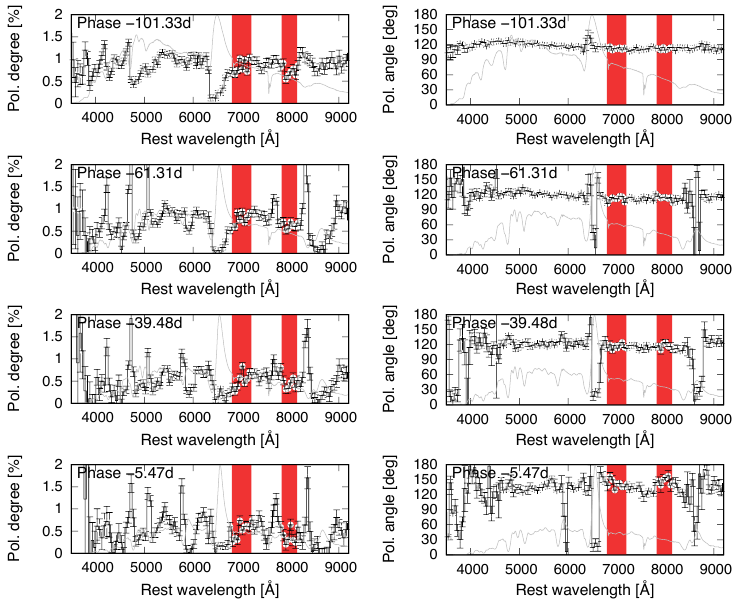}
      \caption{Polarization degree (left) and angle (right) of SN 2021yja. The red hatching shows the adopted wavelength range for the estimation of continuum polarization. The grey lines in the background of each plot are the unbinned flux spectra at the corresponding epochs.
              }
      \label{fig:app1}
   \end{figure*}

\begin{figure*}
            \includegraphics[width=0.5\hsize]{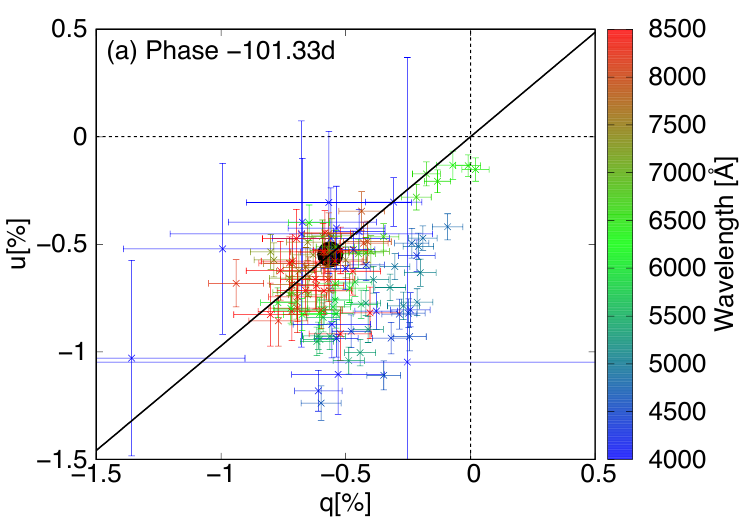}
            \includegraphics[width=0.5\hsize]{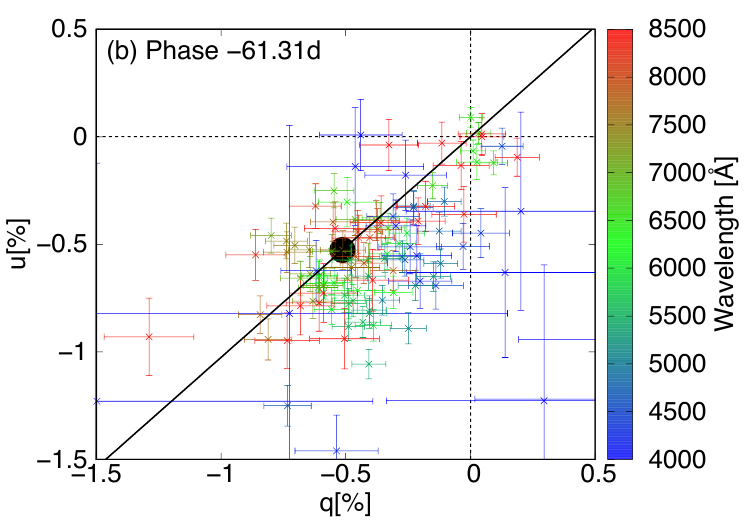}
            \includegraphics[width=0.5\hsize]{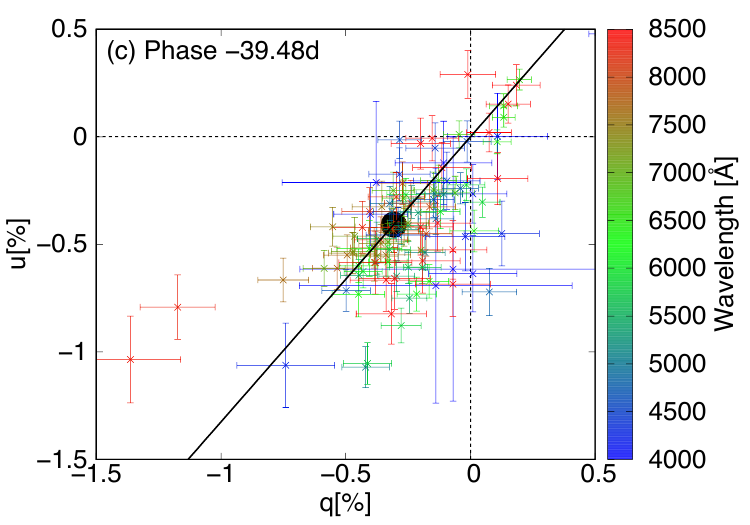}
            \includegraphics[width=0.5\hsize]{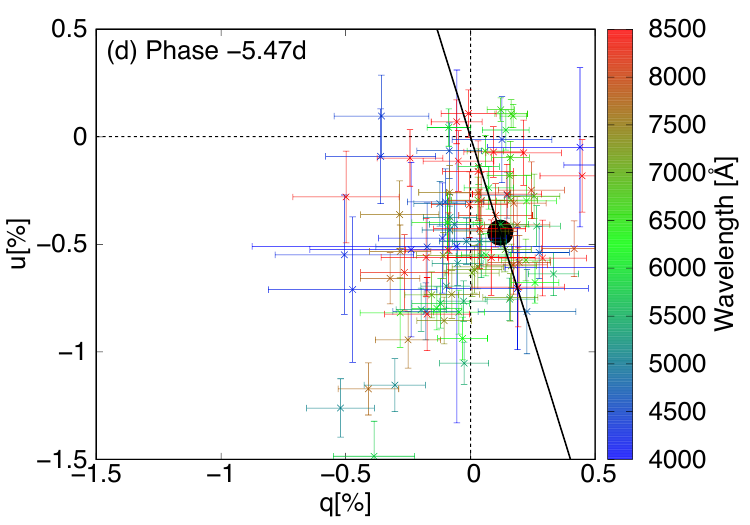}
      \caption{Spectropolarimetric data in the q-u plane. The estimated continuum polarization and its angle are shown with black points and lines, respectively.
              }
      \label{fig:app2}
   \end{figure*}

The observing logs for the spectro- and imaging polarimetry of SN 2021yja are presented in Tables~\ref{tab:spec_pol} and \ref{tab:image_pol}. The full polarization spectra are shown in Figure~\ref{fig:app1} and \ref{fig:app2}.

\begin{table*}
      \caption[]{Log and measurements of the spectropolarimetric observations of SN 2021yja.}
      \label{tab:spec_pol}
      $
         \begin{array}{lccccccc}
            \hline
            \noalign{\smallskip}
            \rm{Date} & \rm{MJD} & \rm{Phase}^{a} & \rm{Days\; from \;explosion}^{b} & \rm{Exp. \;time} & \rm{Pol.\;degree}^{c} & \rm{Pol. \;angle}^{c} & \rm{Telescope}\\
            (\rm{UT}) & (\rm{days}) & (\rm{days}) & (\rm{days}) & (\rm{seconds}) & (\%) & (\rm{degrees}) & \\
            \noalign{\smallskip}
            \hline\hline
            \noalign{\smallskip}
            2021-10-02.24 & 59489.24 & -101.33 & +24.84 & 4 \times 300 & 0.78 (0.00) \pm 0.03 & 112.1 (-) \pm 1.4 & \rm{VLT} \\
            \noalign{\smallskip} \hline \noalign{\smallskip}
            2021-11-11.26 & 59529.26 & -61.31 & +64.86 & 4 \times 300 & 0.74 (0.05) \pm 0.03 & 112.9 (11.1) \pm 3.7 & \rm{VLT} \\
            \noalign{\smallskip} \hline \noalign{\smallskip}
            2021-12-03.09 & 59551.09 & -39.48 & +86.69 & 4 \times 300 & 0.51 (0.29) \pm 0.03 & 116.5 (14.3) \pm 3.0 & \rm{VLT} \\
            \noalign{\smallskip} \hline \noalign{\smallskip}
            2022-01-06.10 & 59585.10 & -5.47 & +120.70 & 4 \times 300 & 0.46 (0.69) \pm 0.04 & 142.5 (4.2) \pm 4.4 & \rm{VLT} \\
            \noalign{\smallskip}
            \hline
         \end{array}
         $
         \begin{minipage}{.88\hsize}
        \smallskip
        Notes. ${}^{a}$Relative to $t_{0}=59590.57$ (MJD), which is the estimated time of the end of the plateau phase. ${}^{b}$Relative to $t=59464.40$ (MJD), which is the estimated explosion time. ${}^{c}$Continuum polarization degree and angle. The values in brackets for the polarization degree and angle refer to the continuum polarization after subtraction of the first component. The errors of the polarization degrees and angles represent the photon shot noise.
        \end{minipage}
   \end{table*}

\begin{table*}
      \caption[]{Log and measurements of the imaging-polarimetric observations of SN 2021yja.}
      \label{tab:image_pol}
      $
         \begin{array}{lcccccccc}
            \hline
            \noalign{\smallskip}
            \rm{Date} & \rm{MJD} & \rm{Phase} & \rm{Days\; from \;explosion} & \rm{Exp. \;time} & \rm{Pol.\;degree} & \rm{Pol. \;angle} & \rm{Filter} & \rm{Telescope}\\
            (\rm{UT}) & (\rm{days}) & (\rm{days}) & (\rm{days}) & (\rm{seconds}) & (\%) & (\rm{degrees}) & & \\
            \noalign{\smallskip}
            \hline\hline
            \noalign{\smallskip}
            \multirow{2}{*}{2021-09-10.18} & \multirow{2}{*}{59467.18} & \multirow{2}{*}{-123.39} & \multirow{2}{*}{+2.78} & 4 \times 40 & 0.08 \pm 0.07 & 110.8 \pm 23.5 & V & \multirow{2}{*}{\rm{NOT}}\\
            & & & & 4 \times 40 & 0.17 \pm 0.07 & 125.0 \pm 11.2 & R & \\
            \noalign{\smallskip} \hline \noalign{\smallskip}
            \multirow{2}{*}{2021-09-16.22} & \multirow{2}{*}{59473.22} & \multirow{2}{*}{-117.35} & \multirow{2}{*}{+8.82} & 4 \times 30 & 0.22 \pm 0.08 & 163.6 \pm 9.9 & V & \multirow{2}{*}{\rm{NOT}}\\
            & & & & 4 \times 30 & 0.73 \pm 0.26 & 124.0 \pm 10.2 & R & \\
            \noalign{\smallskip} \hline \noalign{\smallskip}
            \multirow{2}{*}{2021-09-26.16} & \multirow{2}{*}{59483.16} & \multirow{2}{*}{-107.41} & \multirow{2}{*}{+18.76} & 4 \times 30 & 1.21 \pm 0.10 & 125.9 \pm 2.4 & V & \multirow{2}{*}{\rm{NOT}}\\
            & & & & 4 \times 30 & 0.78 \pm 0.08 & 101.2 \pm 3.0 & R & \\
            \noalign{\smallskip} \hline \noalign{\smallskip}
            \multirow{2}{*}{2021-11-13.01} & \multirow{2}{*}{59531.01} & \multirow{2}{*}{-59.56} & \multirow{2}{*}{+66.61} & 4 \times 30 & 0.68 \pm 0.11 & 104.6 \pm 4.6 & V & \multirow{2}{*}{\rm{NOT}}\\
            & & & & 4 \times 30 & 0.60 \pm 0.09 & 108.1 \pm 4.1 & R & \\
            \noalign{\smallskip} \hline \noalign{\smallskip}
            \multirow{2}{*}{2021-12-23.96} & \multirow{2}{*}{59571.96} & \multirow{2}{*}{-18.61} & \multirow{2}{*}{+107.56} & 4 \times 40 & 0.39 \pm 0.09 & 119.1 \pm 6.4 & V & \multirow{2}{*}{\rm{NOT}}\\
            & & & & 4 \times 30 & 0.60 \pm 0.07 & 135.6 \pm 3.5 & R & \\
            \noalign{\smallskip} \hline \noalign{\smallskip}
            \multirow{2}{*}{2022-02-07.83} & \multirow{2}{*}{59617.83} & \multirow{2}{*}{+27.26} & \multirow{2}{*}{+153.43} & 4 \times 100 & 0.39 \pm 0.29 & 75.1 \pm 21.2 & V & \multirow{2}{*}{\rm{NOT}}\\
            & & & & 4 \times 80 & 0.69 \pm 0.16 & 105.4 \pm 6.6 & R & \\
            \noalign{\smallskip} \hline \noalign{\smallskip}
            2022-03-27.01 & 59665.01 & +74.44 & +200.61 & 4 \times 300 & 0.85 \pm 0.26 & 57.4 \pm 8.9 & \rm{FILT\_815\_13} & \rm{VLT}\\
            \noalign{\smallskip}
            \hline
         \end{array}
         $
   \end{table*}

\section{Photometry} \label{sec:app_photo}

\begin{figure*}
   \centering
            \includegraphics[width=\hsize]{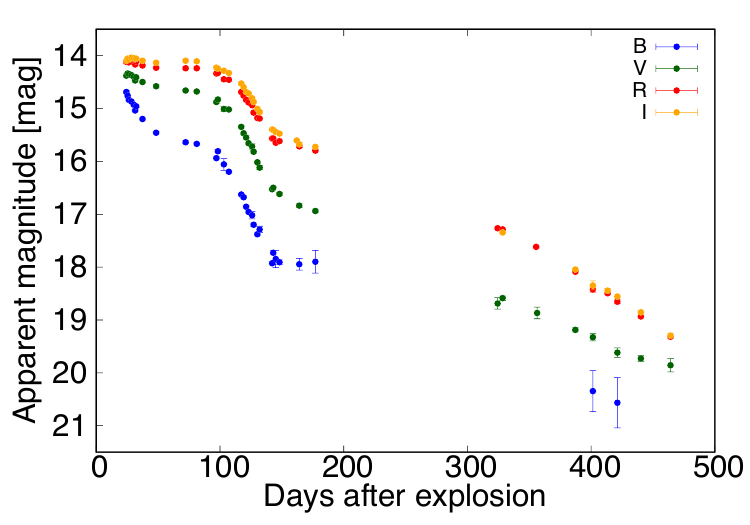}
      \caption{Light curves of SN 2021yja in the optical ($BVRI$) bands. The magnitudes have been corrected for a reddening of $E(B-V)=0.104$
              }
      \label{fig:app3}
   \end{figure*}

\begin{figure*}
   \centering
            \includegraphics[width=\hsize]{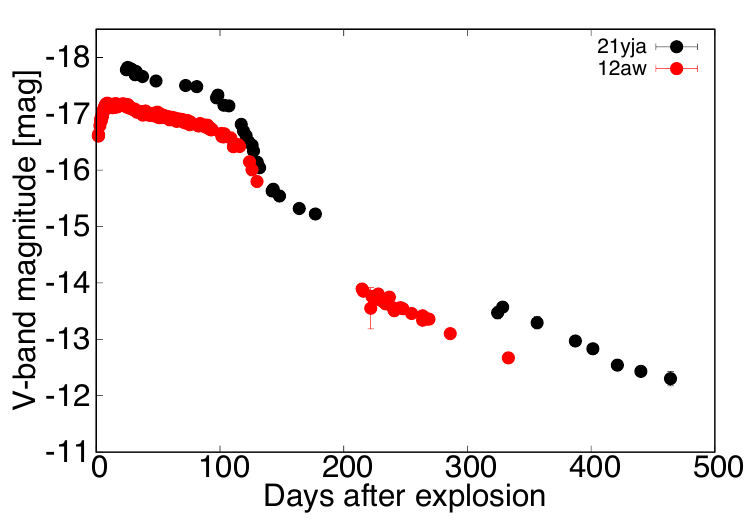}
      \caption{Time evolution of the $V$-band absolute magnitudes of SNe 2021yja and 2012aw. The data of SN~2012aw were taken from \citet[][]{Munari2013,Bose2013,DallOra2014,Brown2014}. The $V$-band extinction for SN~2012aw is assumed to be $A_{V}=0.23$ mag \citep[][]{Bose2013}.
              }
      \label{fig:app4}
   \end{figure*}

We obtained $B$-, $V-$, $R$- and $I$-band photometry of SN 2021yja using the Hiroshima One-shot Wide-field Polarimeter \citep[HOWPol;][]{Kawabata2008} mounted on the 1.5-m  Kanata telescope at the Higashi-Hiroshima Observatory in Japan. 
We performed the point spread function (PSF) photometry using the DAOPHOT package \citep[][]{Stetson1987} as a part of IRAF. Since the host galaxy at the SN position is faint, we did not subtract the template images for the photometry. The magnitudes of the local comparison stars were obtained from the AAVSO Photometric All-Sky Survey \citep[APASS;][]{Henden2012}. We estimated the $R$- and $I$-band magnitudes of the comparison stars from the Sloan magnitudes using the relations in \citet[][]{Jester2005}.
The results of photometry are provided in Table~\ref{tab:phot} and Figures~\ref{fig:app3} and \ref{fig:app4}.

    \begin{table*}
      \caption[]{Log of the photometric observations of SN 2021yja and photometric measurements (Vega magnitudes).}
      \label{tab:phot}
      $
         \begin{array}{cccccc}
            \hline
            \noalign{\smallskip}
            \rm{MJD} & \rm{Days\; from \;explosion} & B & V & R & I\\
            (\rm{days}) & (\rm{days}) & (\rm{mag}) & (\rm{mag}) & (\rm{mag}) & (\rm{mag}) \\
            \noalign{\smallskip}
            \hline\hline
            \noalign{\smallskip}
            59488.75 & +24.35 & 15.11 \pm 0.03 & 14.70 \pm 0.03 & 14.37 \pm 0.03 & 14.27 \pm 0.02\\
	        \noalign{\smallskip} \hline \noalign{\smallskip}
	        59489.72 & +25.32 & 15.18 \pm 0.02 & 14.66 \pm 0.02 & 14.34 \pm 0.02 & 14.23 \pm 0.02\\
	        \noalign{\smallskip} \hline \noalign{\smallskip}
	        59490.78 & +26.38 & 15.26 \pm 0.02 & 14.67 \pm 0.03 & 14.38 \pm 0.02 & 14.25 \pm 0.03\\
            \noalign{\smallskip} \hline \noalign{\smallskip}
            59492.75 & +28.35 & 15.29 \pm 0.02 & 14.69 \pm 0.02 & 14.36 \pm 0.02 & 14.21 \pm 0.02\\
            \noalign{\smallskip} \hline \noalign{\smallskip}
            59494.71 & +30.31 & 15.35 \pm 0.02 & 14.72 \pm 0.02 & 14.35 \pm 0.02 & 14.22 \pm 0.02\\
            \noalign{\smallskip} \hline \noalign{\smallskip}
            59495.79 & +31.39 & 15.46 \pm 0.02 & 14.79 \pm 0.03 & 14.42 \pm 0.02 & 14.24 \pm 0.03\\
            \noalign{\smallskip} \hline \noalign{\smallskip}
            59496.68 & +32.28 & 15.38 \pm 0.02 & 14.73 \pm 0.02 & 14.36 \pm 0.02 & 14.23 \pm 0.03\\
            \noalign{\smallskip} \hline \noalign{\smallskip}
            59501.77 & +37.37 & 15.62 \pm 0.02 & 14.82 \pm 0.02 & 14.44 \pm 0.02 & 14.27 \pm 0.02\\
            \noalign{\smallskip} \hline \noalign{\smallskip}
            59512.74 & +48.34 & 15.88 \pm 0.02 & 14.90 \pm 0.03 & 14.48 \pm 0.02 & 14.31 \pm 0.02\\
            \noalign{\smallskip} \hline \noalign{\smallskip}
            59536.62 & +72.22 & 16.06 \pm 0.03 & 14.98 \pm 0.02 & 14.49 \pm 0.02 & 14.27 \pm 0.02\\
            \noalign{\smallskip} \hline \noalign{\smallskip}
            59545.65 & +81.25 & 16.09 \pm 0.02 & 15.00 \pm 0.02 & 14.49 \pm 0.02 & 14.28 \pm 0.02 \\
            \noalign{\smallskip} \hline \noalign{\smallskip}
            59561.55 & +97.15 & 16.36 \pm 0.02 & 15.20 \pm 0.02 & 14.59 \pm 0.02 & 14.40 \pm 0.02\\
            \noalign{\smallskip} \hline \noalign{\smallskip}
            59562.65 & +98.25 & 16.23 \pm 0.03 & 15.15 \pm 0.03 & 14.58 \pm 0.02 & 14.42 \pm 0.02\\
            \noalign{\smallskip} \hline \noalign{\smallskip}
            59567.56 & +103.16 & 16.48 \pm 0.11 & 15.33 \pm 0.05 & 14.70 \pm 0.04 & 14.46 \pm 0.02\\
            \noalign{\smallskip} \hline \noalign{\smallskip}
            59571.60 & +107.20 & 16.62 \pm 0.03 & 15.34 \pm 0.02 & 14.71 \pm 0.03 & 14.50 \pm 0.02\\
            \noalign{\smallskip} \hline \noalign{\smallskip}
            59581.58 & +117.18 & 17.05 \pm 0.02 & 15.67 \pm 0.02 & 14.94 \pm 0.02 & 14.70 \pm 0.03\\
            \noalign{\smallskip} \hline \noalign{\smallskip}
            59583.48 & +119.08 & 17.10 \pm 0.02 & 15.79 \pm 0.02 & 15.01 \pm 0.02 & 14.77 \pm 0.02\\
            \noalign{\smallskip} \hline \noalign{\smallskip}
            59585.56 & +121.16 & 17.28 \pm 0.02 & 15.87 \pm 0.03 & 15.09 \pm 0.03 & 14.86 \pm 0.02\\
            \noalign{\smallskip} \hline \noalign{\smallskip}
            59587.53 & +123.13 & 17.38 \pm 0.02 & 15.98 \pm 0.02 & 15.14 \pm 0.02 & 14.89 \pm 0.02\\
            \noalign{\smallskip} \hline \noalign{\smallskip}
            59590.41 & +126.01 & 17.44 \pm 0.07 & 16.04 \pm 0.05 & 15.19 \pm 0.02 & 14.98 \pm 0.02\\
            \noalign{\smallskip} \hline \noalign{\smallskip}
            59591.55 & +127.15 & 17.62 \pm 0.03 & 16.14 \pm 0.02 & 15.33 \pm 0.02 & 15.05 \pm 0.02\\
            \noalign{\smallskip} \hline \noalign{\smallskip}
            59594.52 & +130.12 & 17.80 \pm 0.03 & 16.34 \pm 0.02 & 15.43 \pm 0.02 & 15.18 \pm 0.02\\
            \noalign{\smallskip} \hline \noalign{\smallskip}
            59596.52 & +132.12 & 17.71 \pm 0.06 & 16.44 \pm 0.04 & 15.44 \pm 0.02 & 15.24 \pm 0.03\\
            \noalign{\smallskip} \hline \noalign{\smallskip}
            59606.47 & +142.07 & 18.35 \pm 0.03 & 16.85 \pm 0.02 & 15.82 \pm 0.02 & 15.57 \pm 0.02\\
            \noalign{\smallskip} \hline \noalign{\smallskip}
            59607.46 & +143.06 & 18.15 \pm 0.04 & 16.82 \pm 0.02 & 15.82 \pm 0.02 & 15.58 \pm 0.02\\
            \noalign{\smallskip} \hline \noalign{\smallskip}
            59609.46 & +145.06 & 18.27 \pm 0.16 & -              & 15.90 \pm 0.02 & 15.62 \pm 0.03\\
            \noalign{\smallskip} \hline \noalign{\smallskip}
            59612.48 & +148.08 & 18.33 \pm 0.05 & 16.94 \pm 0.03 & 15.87 \pm 0.02 & 15.65 \pm 0.02\\
            \noalign{\smallskip} \hline \noalign{\smallskip}
            59626.49 & +162.09 & -              & -              & -              & 15.78 \pm 0.02\\
            \noalign{\smallskip} \hline \noalign{\smallskip}
            59628.44 & +164.04 & 18.37 \pm 0.11 & 17.16 \pm 0.03 & 15.97 \pm 0.02 & 15.85 \pm 0.03\\
            \noalign{\smallskip} \hline \noalign{\smallskip}
            59641.44 & +177.04 & 18.32 \pm 0.21 & 17.26 \pm 0.03 & 16.05 \pm 0.02 & 15.90 \pm 0.02\\
            \noalign{\smallskip} \hline \noalign{\smallskip}
            59788.80 & +324.40 & -              & 19.01 \pm 0.11 & 17.52 \pm 0.02 & -             \\
            \noalign{\smallskip} \hline \noalign{\smallskip}
            59792.79 & +328.39 & -              & 18.91 \pm 0.04 & 17.54 \pm 0.02 & 17.52 \pm 0.02\\
            \noalign{\smallskip} \hline \noalign{\smallskip}
            59819.82 & +355.42 & -              & -              & 17.87 \pm 0.02 & -             \\
            \noalign{\smallskip} \hline \noalign{\smallskip}
            59820.78 & +356.38 & -              & 19.19 \pm 0.11 & -              & -             \\
            \noalign{\smallskip} \hline \noalign{\smallskip}
            59851.71 & +387.31 & -              & 19.51 \pm 0.04 & 18.34 \pm 0.02 & 18.22 \pm 0.02\\
            \noalign{\smallskip} \hline \noalign{\smallskip}
            59865.72 & +401.32 & 20.77 \pm 0.39 & 19.65 \pm 0.07 & 18.68 \pm 0.05 & 18.52 \pm 0.08\\
            \noalign{\smallskip} \hline \noalign{\smallskip}
            59877.74 & +413.34 & -              & -              & 18.75 \pm 0.02 & 18.62 \pm 0.04\\
            \noalign{\smallskip} \hline \noalign{\smallskip}
            59885.58 & +421.18 & 20.99 \pm 0.48 & 19.94 \pm 0.09 & 18.91 \pm 0.04 & 18.73 \pm 0.03\\
            \noalign{\smallskip} \hline \noalign{\smallskip}
            59904.58 & +440.18 & -              & 20.05 \pm 0.06 & 19.19 \pm 0.03 & 19.03 \pm 0.03\\
            \noalign{\smallskip} \hline \noalign{\smallskip}
            59928.57 & +464.17 & -              & 20.18 \pm 0.13 & 19.57 \pm 0.03 & 19.47 \pm 0.05\\
            \noalign{\smallskip}
            \hline
         \end{array}
         $
   \end{table*}

\section{Spectroscopy} \label{sec:app_spec}

In addition to the spectra obtained from the spectropolarimetric observations with the FORS2/VLT (see Section~\ref{sec:observation}), we obtained optical spectra of SN 2021yja using the Kyoto Okayama Optical Low-dispersion Spectrograph with optical-fiber Integral Field Unit \citep[KOOLS-IFU;][]{Matsubayashi2019} mounted on the Seimei telescope \citep[][]{Kurita2020} at the Okayama Observatory of Kyoto University and the Alhambra Faint Object Spectrograph and Camera (ALFOSC)\footnote{\url{https://www.not.iac.es/instruments/alfosc/}} mounted on the 2.56-m Nordic Optical Telescope (NOT)\footnote{\url{https://www.not.iac.es/}} at the Roque de los Muchachos Observatory. The log of spectroscopic observations are provided in Table~\ref{tab:spec}.

\begin{figure*}
   \centering
            \includegraphics[width=\hsize]{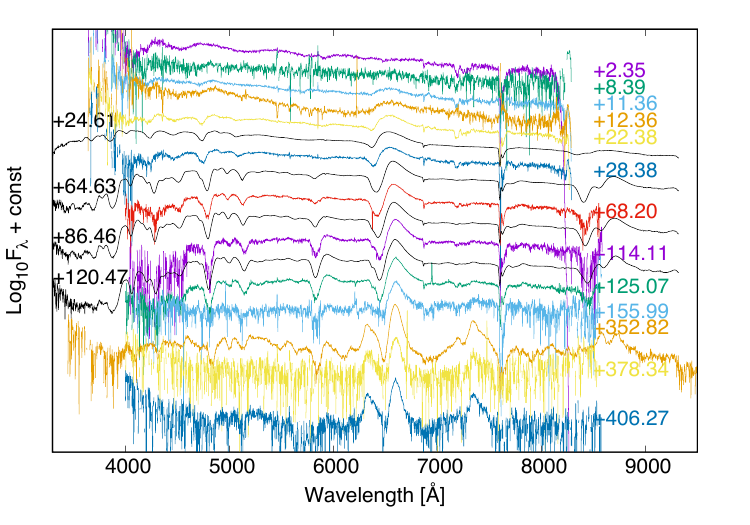}
      \caption{Spectral evolution of SN~2021yja. The numbers show days from the explosion.
              }
      \label{fig:app5}
   \end{figure*}

\begin{figure*}
   \centering
            \includegraphics[width=\hsize]{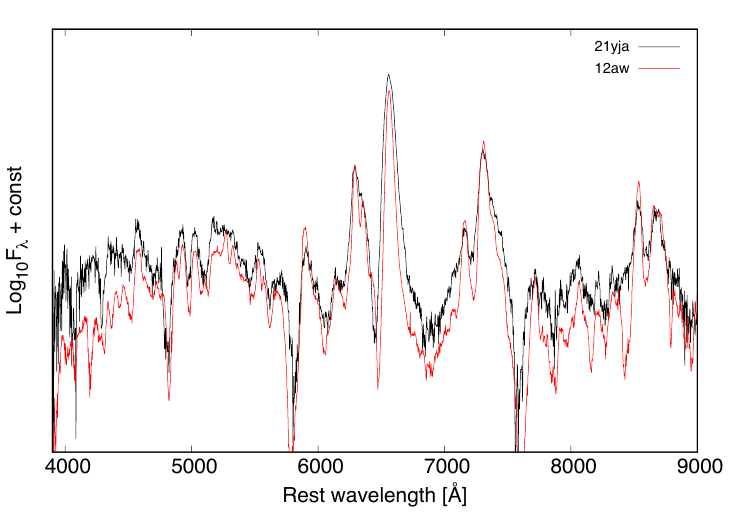}
      \caption{Same as Figure~\ref{fig:fig4}, but for the full wavelength range.
              }
      \label{fig:app6}
   \end{figure*}

For the KOOLS-IFU/Seimei, we used the VPH-blue grism, giving a wavelength coverage of 4100–8900 {\AA} and a spectral resolution of $R = \lambda/\Delta\lambda \sim500$. The data reduction was performed following the standard procedures that are developed for KOOLS-IFU data\footnote{\url{http://www.o.kwasan.kyoto-u.ac.jp/inst/p-kools/reduction-201806/index.html}}, using the IRAF. For the ALFOSC/NOT, we used Grism 4, giving a wavelength coverage of 3200–9600 {\AA} and a spectral resolution of $\sim 360$. The spectrum was reduced with the alfoscgui pipeline\footnote{FOSCGUI is a graphical user interface aimed at extracting SN spectroscopy and photometry obtained with FOSC-like instruments. It was developed by E. Cappellaro. A package description can be found at \url{https://sngroup.oapd.inaf.it/foscgui.html}}. The procedures include overscan, bias, and flat-field corrections, as well as removing cosmic-ray artefacts. Extraction of a one-dimensional spectrum and sky subtraction were performed. Wavelength calibration was performed by comparison with arc lamps. The spectra were flux-calibrated against a sensitivity function derived from a standard star observed on the same night.
The reduced spectra are shown in Figures~\ref{fig:app5} and \ref{fig:app6}.

\begin{table*}
      \caption[]{Log of the spectroscopic observations of SN 2021yja.}
      \label{tab:spec}
      $
         \begin{array}{lccccc}
            \hline
            \noalign{\smallskip}
            \rm{Date} & \rm{MJD} & \rm{Phase} & \rm{Days\; from \;explosion} & \rm{Exp. \;time} & \rm{Instrument/Telescope}\\
            (\rm{UT}) & (\rm{days}) & (\rm{days}) & (\rm{days}) & (\rm{seconds}) & \\
            \noalign{\smallskip}
            \hline\hline
            \noalign{\smallskip}
            2021-09-09.75 & 59466.75 & -123.82 & +2.35 & 300 \times 3 & \rm{KOOLS-IFU/Seimei} \\
            \noalign{\smallskip} \hline \noalign{\smallskip}
            2021-09-15.79 & 59472.79 & -117.78 & +8.39 & 600 \times 3  & \rm{KOOLS-IFU/Seimei} \\
            \noalign{\smallskip} \hline \noalign{\smallskip}
            2021-09-18.76 & 59475.76 & -114.81 & +11.36 & 300 \times 3 & \rm{KOOLS-IFU/Seimei} \\
            \noalign{\smallskip} \hline \noalign{\smallskip}
            2021-09-19.76 & 59476.76 & -113.81 & +12.36 & 600 \times 3 & \rm{KOOLS-IFU/Seimei} \\
            \noalign{\smallskip} \hline \noalign{\smallskip}
            2021-09-29.78 & 59486.78 & -103.79 & +22.38 & 900 \times 3 & \rm{KOOLS-IFU/Seimei} \\
            \noalign{\smallskip} \hline \noalign{\smallskip}
            2021-10-02.24 & 59489.24 & -101.33 & +24.84 & 300 \times 4 & \rm{FORS2/VLT} \\
            \noalign{\smallskip} \hline \noalign{\smallskip}
            2021-10-05.78 & 59492.78 & -97.79 & +28.38 & 300 \times 3 & \rm{KOOLS-IFU/Seimei} \\
            \noalign{\smallskip} \hline \noalign{\smallskip}
            2021-11-11.26 & 59529.26 & -61.31 & +64.86 & 300 \times 4 & \rm{FORS2/VLT} \\
            \noalign{\smallskip} \hline \noalign{\smallskip}
            2021-11-14.60 & 59532.60 & -59.97 & +68.20 & 600 \times 3 & \rm{KOOLS-IFU/Seimei} \\
            \noalign{\smallskip} \hline \noalign{\smallskip}
            2021-12-03.09 & 59551.09 & -39.48 & +86.69 & 300 \times 4 & \rm{FORS2/VLT} \\
            \noalign{\smallskip} \hline \noalign{\smallskip}
            2021-12-30.51 & 59578.51 & -12.06 & +114.11 & 1200 \times 2 & \rm{KOOLS-IFU/Seimei} \\
            \noalign{\smallskip} \hline \noalign{\smallskip}
            2022-01-06.10 & 59585.10 & -5.47 & +120.70 & 300 \times 4 & \rm{FORS2/VLT} \\
            \noalign{\smallskip} \hline \noalign{\smallskip}
            2022-01-10.47 & 59589.47 & -1.1 & +125.07 & 1200 \times 1 & \rm{KOOLS-IFU/Seimei} \\
            \noalign{\smallskip} \hline \noalign{\smallskip}
            2022-02-10.39 & 59620.39 & +29.82 & +155.99 & 600 \times 2 & \rm{KOOLS-IFU/Seimei} \\
            \noalign{\smallskip} \hline \noalign{\smallskip}
            2022-08-26.22 & 59817.22 & +226.65 & +352.82 & 1800 \times 1 & \rm{ALFOSC/NOT} \\
            \noalign{\smallskip} \hline \noalign{\smallskip}
            2022-09-21.74 & 59842.74 & +252.17 & +378.34 & 900 \times 1 & \rm{KOOLS-IFU/Seimei} \\
            \noalign{\smallskip} \hline \noalign{\smallskip}
            2022-10-18.67 & 59870.67 & +280.10 & +406.27 & 900 \times 3 & \rm{KOOLS-IFU/Seimei} \\
            \noalign{\smallskip}
            \hline
         \end{array}
         $
   \end{table*}

\end{document}